\documentclass{article}
\usepackage{times}
\usepackage{latexsym}
\usepackage{amsmath, amssymb}
\usepackage{graphicx}
\usepackage{hyperref}
\usepackage[numbers]{natbib}
\usepackage{authblk}

\title{An End-to-End Homomorphically Encrypted Neural Network$^{\dagger}$ \vspace{10mm}}

\author{
    \begin{minipage}{0.3\textwidth} \centering
        \textbf{\small Marcos Florencio*} \\  
         \vspace{-0.5em}
        \scriptsize INTELI - Institute of Technology and Leadership \\  
        \texttt{\tiny marcos.florencio@sou.inteli.edu.br}
    \end{minipage}
    \hfill
    \begin{minipage}{0.3\textwidth} \centering
        \textbf{\small Luiz Alencar*} \\  
         \vspace{-0.5em}
        \scriptsize INTELI - Institute of Technology and Leadership \\  
        \texttt{\tiny luiz.alencar@sou.inteli.edu.br}
    \end{minipage}
    \hfill
    \begin{minipage}{0.3\textwidth} \centering
        \textbf{\small Bianca Lima*} \\ 
         \vspace{-0.5em}
        \scriptsize INTELI - Institute of Technology and Leadership \\  
        \texttt{\tiny bianca.lima@sou.inteli.edu.br}
    \end{minipage}
}

\date{}

\begin{document}
\maketitle

\begin{abstract}
Every commercially available, state-of-the-art neural network consume plain input data, which is a well-known privacy concern. We propose a new architecture based on homomorphic encryption, which allows the neural network to operate on encrypted data. We show that Homomorphic Neural Networks (HNN) can achieve full privacy and security while maintaining levels of accuracy comparable to plain neural networks. We also introduce a new layer, the Differentiable Soft-Argmax, which allows the calibration of output logits in the encrypted domain, raising the entropy of the activation parameters, thus improving the security of the model, while keeping the overall noise below the acceptable noise budget. Experiments were conducted using the Stanford Sentiment Treebank (SST-2) corpora on the DistilBERT base uncased finetuned SST-2 English sentiment analysis model, and the results show that the HNN model can achieve up to 82.5\% of the accuracy of the plain model while maintaining full privacy and security.
\end{abstract}

\vspace{25mm}

{\scriptsize$^{\dagger}$ 
This is an update of the version published in February 22, 2025, due to the inclusion of patent number BR 10 2025 003497 2.}

{\scriptsize*Equal contribution. Listing order is random. Florencio proposed the use of homomorphic encryption in neural networks and was involved in all aspects of this work. He also designed the first version of the Differentiable Soft-Argmax layer. Alencar implemented the final versions of the training pipelines and the Differentiable Soft-Argmax layer, and was responsible for the experiments. Lima was in charge of putting together the final version of the paper and the experiments and also spent countless hours developing an interactive demo for the model, available at https://www.epiphyte-2.com.}
\vspace{20mm}

\section{Introduction}

At the same time that neural networks are considered "black-boxes," due to their inherent internal complexity, the need to relinquish privacy to use state-of-the-art neural networks is a well-known issue, especially for companies that deal with sensitive data \cite{buhrmester2021analysis,savage2022breaking}. In these models, the input of plain text is converted into a vector of features that the model uses to perform inference. If the input could be encrypted, the model could be used without the need to share the plain data with the model owner, providing full privacy and security.

We adapt the encryption scheme proposed by \cite{cheon2017homomorphic}, which is designed to perform computation on real numbers, to encrypt input data, plus some acceptable level of noise, and then calibrate the model's weights in such a way that the forward pass of the neural network can be performed in the encrypted domain. In the same fashion, the outputs of the model are calibrated so that only the owner of the private key can decrypt the output. The main challenge of this approach is the noise that is introduced in the encrypted domain, which can be mitigated by using a noise budget, which is a parameter that limits the maximum amount of noise. While the base scheme requires specific Add and Mult algorithms to perform operations, they are performed internally in the context of neural networks and the noise budget is controlled by the model's architecture and the training process during backpropagation.

It is also commom practice in neural network architectures that the output of the model is a vector of raw logits, upon which the softmax function is applied to calculate the loss function and to perform inference\cite{he2016deep,oord2016wavenet,vaswani2017attention, srivastavan2014dropout}. It is our understanding that a privacy-preserving neural network should maintain the output logits in the encrypted domain, ideally introducing new noise. To address this issue, we propose a new layer, the Differentiable Soft-Argmax, which is a differentiable version of the argmax function. It calibrates the logits in the encrypted domain, raising the entropy of the activation parameters, thus improving the security of the model while keeping the overall noise below the acceptable noise budget.

\vspace{10mm}

\section{Background}

Homomorphically encrypted neural networks is a relatively new field of research, with the first significant work being published in 2016 \cite{gilad2016cryptonets}. It demonstrated how a network can execute over encrypted data and introduced optimizations to make inference feasible. The encryption scheme chosen was YASHE', described in \cite{bos2013improved}, which does not support floating-point numbers and allows operations on encrypted data if, and only if, the complexity of the arithmetic circuit is known in advance. 

Both of these limitations were addressed by \cite{cheon2017homomorphic}, which introduced a new encryption scheme that allows operations on encrypted data without the need of knowing the complexity of the arithmetic circuit in advance. This scheme is known as HEANN or CKKS, and it is the one we use in this work. Another fundamental difference is the optizations proposed by \cite{gilad2016cryptonets}, such as replacing activation functions by polynomials, which end up tampering with the model's architecture, making their approach less generalizable. We rely solely on backpropagation to perform all operations internally while keeping the noise budget controlled. This is acceptable because, even though the weights are known, the underlying embeddings are unknown, due to the noise added during encryption.

Since then, several works have been published, such as \cite{lee2023precise}, which implements word-wise encryption and replaces activation functions by precise polynomial approximation techniques, \cite{legiest2023neural}, which proposes quantization as a way to allow more efficient computations, and \cite{montero2024neural}, which focuses on using homomorphic encryption as a solution to build training approachs on encrypted data combined with multi-party computation. 

As best we can determine, we believe that this is the first work that proposes an end-to-end homomorphically encrypted neural network that relies solely on encryption and backpropagation to train the model. We also believe that this is the first work that formally introduces a new layer, which we call Differentiable Soft-Argmax, that can be used to calibrate the output logits in the encrypted domain.


\vspace{10mm}

\section{Homomorphic encryption}

The objective of using homomorphic encryption \(\varepsilon\) is that, given an algorithm \(\text{Evaluate}_\varepsilon\) and a valid public key \text{pk}, a computation can be performed on any circuit \(\mathcal{C}\) (i.e., a collection of logic gates used to compute a function) and any cyphertexts \(\psi_i \leftarrow \text{Encrypt}_{\varepsilon}(\text{pk}, \pi_i)\), such that the output of the computation is an encryption of $\psi \leftarrow \text{Evaluate}_{\varepsilon}(\text{pk}, \mathcal{C}, \psi_1, \dots, \psi_t)$
and that \(\text{Decrypt}_{\varepsilon}(\text{sk}, \psi) = \mathcal{C}(\pi_1, \dots, \pi_t)\) for a valid secret key \(\text{sk}\), as first described by \cite{rivest1978data}. The first viable construction of a fully homomorphic encryption scheme was proposed by \cite{gentry2009fully} and may be broken down into three major conceptual steps:

\begin{itemize}
    \item a \emph{"somewhat homomorphic"} scheme, supporting evaluation of low-degree polynomials on encrypted data;
    \item a \emph{"squashing"} decryption mechanism, responsible for expressing outputs as low-degree polynomials supported by the scheme, and;
    \item a \emph{"bootstrapping"} transformation, a self-referential property that makes the depth of the decryption circuit shallower than what the scheme can handle.
\end{itemize}

Gentry's main insight was to use bootstrapping in order to obtain a scheme that could evaluate polynomials of high-enough degrees while keeping the decryption procedure expressed as polynomials of low-enough degrees, so that the degrees of the polynomials evaluated by the scheme could surpass the degrees of the polynomials decrypted by the scheme \cite{gentry2011implementing}. This scheme used \textbf{"ideal lattices"}, corresponding to ideals in polynomial rings, to perform homomorphic operations. The reason for this is that ideal lattices inherit natural addition and multiplication properties from the ring, which allows for homomorphic operations to be performed on encrypted data. We defer the discussion on ideal lattices to the next section, since these operations are performed in the context of neural networks in our case.

Several implementations of homomorphic encryption schemes have been developed since, including DGHV \cite{van2010fully}, BGV \cite{brakerski2014leveled} and BFV \cite{fan2012somewhat}, with varying levels of security and efficiency. In 2016, Cheon \emph{et al.} introduced HEANN (a.k.a. CKKS), which is a variant of the BFV scheme designed for computations on real numbers. This scheme is also particularly well-suited for neural network evaluation, where each layer is composed of several matrix--vector multiplications and activation functions that can have their inputs approximated by encrypted floating-point polynomials.

Following the methodology adopted by \cite{gentry2011implementing}, we derive a scheme from HEANN capable of performing inference on encrypted data when applied to neural networks. The main idea is to encode a message \(\pi\) into a plaintext polynomial \(m\) and encrypt it using a public key \(\text{pk}\) to obtain a ciphertext \(\text{Enc}(m, \nu)\). In this scheme, \(\nu\) stands for the \emph{noise parameter} attached to each polynomial \(m\) such that the noise is less than some threshold \(\nu \gg \nu\), as proposed by \cite{gentry2009fully}. This \emph{noise budget} is a critical parameter in the scheme and is inserted to guarantee the security of hardness assumptions such as LWE \cite{regev2009lattices} and RLWE \cite{lyubashevsky2010ideal}. In this sense, it determines the security level of the encryption and the scheme is designed to ensure that the noise budget is not exceeded at any point during the computation.

\[
(m, \nu) \xleftarrow{\text{Encode}} \pi
\]

As in \cite{cheon2017homomorphic}, “it is inevitable to encrypt a vector of multiple plaintexts in a single ciphertext for efficient homomorphic computation” (p.3), being the plaintext space usually a cyclotomic ring \(\mathbb{Z}_t[X]/(\Phi_M(X))\) of a finite characteristic. Thus, after being encoded by a native plaintext space, i.e., the canonical embedding map, the plaintext polynomial \((m, \nu)\) is encrypted using the public key \(\text{pk}\) to obtain the ciphertext \(\text{Enc}(m, \nu)\). We use a \emph{probabilistic public key encryption} scheme so that the encryption function is randomized, such that the same plaintext polynomial \(m\) can be encrypted into different ciphertexts \(\text{Enc}(m, \nu)\) for different instances of the model, and encryption can be performed by anyone with access to the public key \(\text{pk}\), enabling the model to consume data encrypted by any party.

\[
\text{Enc}(m, \nu) \xleftarrow{\text{pk}} (m, \nu)
\]

The encrypted data can then be used to perform computations as it is fed into the neural network. Considering the approximate arithmetic nature of our base scheme, the \emph{noise parameter} is treated as part of error during approximate downstream matrix operations. In this sense, consider that the encryption \(\mathbf{c}\) of message \(\pi\) by the public key \(\text{pk}\) will have a structure of the form \(\langle \mathbf{c}, \text{pk} \rangle = m + \nu\), where \(\nu\) must be large enough to mask any significant features of message \(\pi\) while not exceeding the noise budget \(\nu\). Therefore, for a noise budget of \(0 \leq \nu \leq \nu\), each element of the output vector of the base neural network contains an embedded Gaussian distributed error \(\nu'\) such that \(\lvert \nu' \rvert \leq \nu\).

\[
\text{Enc}(m', \nu') \xleftarrow{f(\cdot)} \text{Enc}(m, \nu)
\]

The encrypted output can then be decrypted using a secret key \(\text{sk}\) to obtain the plaintext polynomial \(m'\). For each instance of the network, the decryption function is deterministic, such that the same ciphertext \(\text{Enc}(m', \nu')\) will always decrypt to the same plaintext polynomial \(m'\). The decryption function is also designed to ensure that the noise parameter \(\nu'\) is removed from the decrypted plaintext polynomial \(m'\) during decryption.

\[
m' \xleftarrow{\text{sk}} \text{Enc}(m', \nu')
\]

Finally, the plaintext polynomial \(m'\) is decoded to obtain the message \(\pi'\), which represents the output prediction of the neural network for the input message \(\pi\).

\[
\pi' \xleftarrow{\text{Decode}} m'
\]

The HEANN scheme consists of five algorithms \((KeyGen, Enc, Dec, Add, Mult)\), while we use here only the first three algorithms \((KeyGen, Enc, Dec)\), since the neural network will be performing additions and multiplications internally. The \(KeyGen\) algorithm generates the public and secret keys, the \(Enc\) algorithm encrypts an encoded input plaintext polynomial and the \(Dec\) algorithm decrypts an output ciphertext into an encoded output plaintext polynomial. The parameters adopted are based on the original HEANN scheme and follow the notation and definitions established in the (2018, Standard):

\begin{itemize}
    \item \(\text{ParamGen}(\lambda, \text{PT}, \text{K}, \text{B}) \rightarrow \text{Params}\): the parameter generation algorithm, used to instantiate various parameters used in the core HE algorithms.
    \begin{itemize}
        \item \(\lambda\) is the desired security level parameter, e.g., 128-bit security (\(\lambda = 128\)) or 256-bit security (\(\lambda = 256\)).
        \item \(\text{PT}\) is the underlying plaintext space of Gaussian distributed numbers mapping the message \(\pi\) to a polynomial \(m\).
        \item \(\text{K}\) is the number of dimensions of the vectors to be encrypted \((\text{V}_1, \ldots, \text{V}_k)\).
        \item \(\text{B}\) is an indirect parameter that determines the key sizes, ciphertext sizes, and complexity of the evaluation procedures.
    \end{itemize}
    
    \item \(KeyGen(\lambda) \rightarrow (\text{pk}, \text{sk})\): Generates a public key \(\text{pk}\) and a secret key \(\text{sk}\), where \(\lambda\) is the security parameter.
    \begin{itemize}
        \item Given the security parameter \(\lambda\), choose a random integer \(h\), a byte length \(\text{P}\) for the key, and fix the number of iterations \(\text{I}\) to ensure sufficient computational cost.
        \item Derive the key as: \(k \leftarrow \mathcal{f}(\lambda, h, \text{P}, \text{I})\) applying the key derivation function iteratively.
    \end{itemize}
    
    \item \(Enc(\text{pk}, m) \rightarrow \text{Enc}(m, \nu)\): Encrypts a plaintext polynomial \(m\) using the public key \(\text{pk}\) to obtain a ciphertext \(\text{Enc}(m, \nu)\).
    \begin{itemize}
        \item for an \(n\)-dimensional vector \(m = (m_1, \ldots, m_n)\) where \(m_i \in \text{PT}\), compute the vector \(m, v = (m_1, v_1, \ldots, m_n, v_n)\) where \(n = \text{K}\) and \(i_{\text{len}} = \text{B}\) using the public key \(\text{pk}\) to obtain a ciphertext \(\text{Enc}(m, \nu)\).
    \end{itemize}
    
    \item \(Dec(\text{sk}, \text{Enc}(m', \nu')) \rightarrow m'\): Decrypts a ciphertext \(\text{Enc}(m', \nu')\) using the secret key \(\text{sk}\) to obtain a plaintext polynomial \(m'\).
\end{itemize}

Similar to Cheon \emph{et al.} (2016, p.9), we do not use a separate plaintext space from an inserted error, so the output \(m' = f(m, \nu)\) is slightly different from the output for the original message \(m\), but the error is considered an approximate value for approximate computations and eventually corrected during activation functions embedded in the neural network. We also do not use Galois keys and bootstrapping, as we are not performing any rotations on the ciphertexts and we do not need to perform dimensional reductions in the encrypted domain.

\section{Neural network architecture}

Neural network interpretability is a long known and mostly unsolved problem in the field of artificial intelligence \cite{zeiler2014visualizing,karpathy2015visualizing}, sometimes being referred to as \textit{black boxes} \cite{fong2017interpretable}. Therefore, input/output privacy-preserving neural networks allow the use of models on sensitive data without compromising the privacy of the data.

\begin{figure}[h]
    \centering
    \includegraphics[width=1\textwidth]{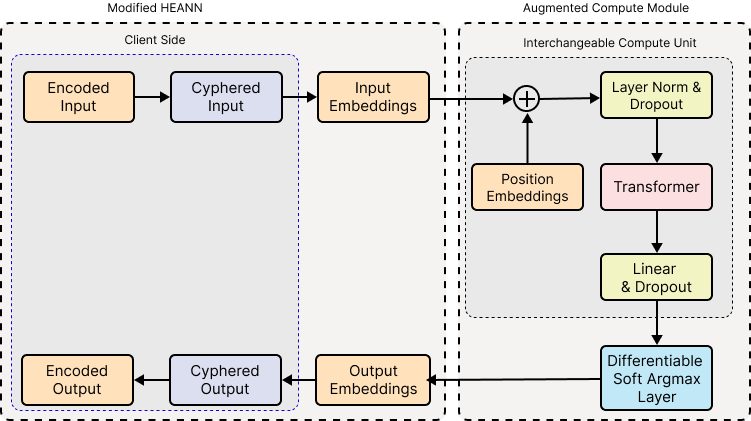}
    \caption{Proposed Architecture}
\end{figure}

In general, neural networks can be considered a regression function that outputs data based on elaborate relationships between high-dimensional input data. As observed by \cite{marcolla2022survey}, privacy-preserving neural networks tend to suffer from high computational complexity and low efficiencies, as the computational complexity of training neural networks is generally high. By avoiding the need of training on encrypted data, we can reduce the computational complexity of the model and improve its efficiency.

We propose an architecture (Figure 1) that leverages a modified version of \texttt{HEANN} to encrypt encoded inputs and decrypt outputs, while introducing a new layer that performs the activation of raw logits in the encrypted domain, which we call the \textit{Differentiable Soft Argmax layer}. This layer is designed to approximate a soft argmax function in the encrypted domain, allowing the neural network to perform prediction without leaking any intermediate values. By being differentiable, it also allows the use of backpropagation to calibrate the logits before ingestion by the softmax function, adding additional noise $\nu$ to the intermediate output of the base model, while effectively keeping it below the \emph{noise budget} $\nu \gg \nu$.

\subsection{Rings, Ideals and Lattices}

The requirement of using \textbf{ideal lattices} comes from the necessity of using encryption schemes whose decryption algorithms have low circuit complexity, generally represented by matrix-vector multiplication, with the important caveat that ``code-based constructions would also represent an interesting possibility'' \cite{gentry2009fully}.

A ring is a set $R$ closed under two binary operations $+$ and $\times$ and with an additive identity $0$ and a multiplicative identity $1$. An ideal $I$ of a ring $R$ is a subset $I \subseteq R$ such that $\sum_{j=1}^{t} i_j + i'_j \in I 
\quad \text{and} \quad
\sum_{j=1}^{t} i_j \times r_j \in I $
for $i_1, \dots, i_t \in I$ and $r_1, \dots, r_t \in R$. As proposed by Gentry, the public key $\text{pk}$ contains an ideal $I$ and a plaintext space $\mathcal{P}$, consisting of cosets of the ideal $I$ in the ring $R$, while the secret key $\text{sk}$ corresponds to a short ideal $I$ in $R$. To encrypt $\pi \in \mathcal{P}$, the encrypter performs 
$\psi \xleftarrow{R} \pi + I,$
where $I$ represents the noise parameter. The decrypter then performs 
$\pi \xleftarrow{R} \psi - I$
to decrypt the ciphertext $\psi$. To perform add and multiply operations on the encrypted data, ring homomorphisms are used on the plaintext space $\mathcal{P}$:
\[
\text{Add}(\text{pk}, \psi_1, \psi_2) = \psi_1 + \psi_2 \in (\pi_1 + \pi_2) + I,
\]
\[
\text{Mult}(\text{pk}, \psi_1, \psi_2) = \psi_1 \times \psi_2 \in (\pi_1 \times \pi_2) + I.
\]

These ciphertexts are essentially ``noisy'' lattice vectors, or ring elements, and decrypting them requires knowledge of the basis for a particular lattice \cite{ajtai1996generating}. An \emph{ideal lattice} is formed by embedding a ring ideal into a real $n$-dimensional coordinate space $\mathbb{R}^n$ used to represent the elements of an ideal as vectors \cite{lyubashevsky2010ideal}. Because $\mathbb{R}$ is discrete and has finite rank $n$, the images of its elements under the embedding form a lattice. Ideal lattices allow ciphertext operations to be performed efficiently using polynomial arithmetic, e.g., Fast Fourier Transform-based multiplication \cite{regev2009lattices}.

\begin{figure}[h]
    \centering
    \includegraphics[width=0.6\textwidth]{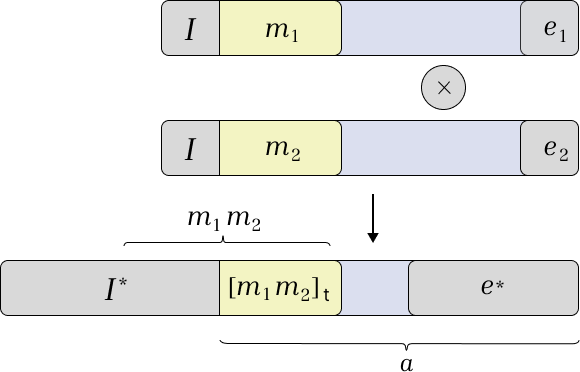}
    \caption{Ideals, Source: Cheon, 2016}
\end{figure}

\noindent
Consider plaintext messages $m_1$ and $m_2$ as in Figure 2. After encryption using the public key $\text{pk}$, these messages become vectors $\psi_1$ and $\psi_2$, each containing the underlying plaintexts, the ideals $\text{I}_1$ and $\text{I}_2$ used to obtain the cyphertexts, and some error $e_1$ and $e_2$ stemming from the basis used to derive the secret key $\text{sk}$. A bit-wise homomorphic operation $\text{Mult}(\text{pk}, \psi_1, \psi_2)$ can then be performed on the cyphertexts to obtain the cyphertext $\psi_1 \times \psi_2$, which contains the product of the plaintexts $m_1 \times m_2$, the ideals $\text{I}_1$ and $\text{I}_2$, and the errors $e_1$ and $e_2$.

\subsection{Modified HEANN}

HEANN supports approximate addition and multiplication of encrypted messages by truncating a ciphertext into a smaller modulus, which leads to rounding of plaintext and also adds noise to the message. The main purpose of the noise is for security reasons and, given the nature of the scheme, it ends up being reduced during computation due to rescaling. As the authors explain (p. 6), ``the most important feature of our scheme is the rounding operation of plaintexts.'' The output of this scheme, therefore, is an approximate value with a predetermined precision.

As per the scheme definition, given the set of complex numbers
\[
\mathbb{H} 
= \Bigl\{ (z_j)_{j \in \mathbb{Z}_M^*} \colon z_{-j} = \overline{z_j}, 
\ \forall j \in \mathbb{Z}_M^* \Bigr\}
\subset \mathbb{C}^{\Phi(M)},
\]
and the subgroup $T$ of the multiplicative group $\mathbb{Z}_M^*$ satisfying $\mathbb{Z}_M^*/T = \{\pm 1\}$, the input of the scheme is a plaintext polynomial constituted by elements of a cyclotomic ring 
\[
R 
= \mathbb{Z}_t[X]/\bigl(\Phi_M(X)\bigr),
\]
which is mapped to a vector of complex numbers via an embedding map represented by a ring homomorphism. This transformation enables the preservation of the precision of the plaintext after encoding. The decoding procedure is almost the inverse of the encoding, except for the addition of a round-off algorithm for discretization. An important characteristic of HEANN that makes it suitable for neural networks is that the bit size of ciphertext modulus does not grow exponentially in relation to the circuit being evaluated, which allows for the evaluation of deep neural networks without the need for bootstrapping.

In our version, based on Gentry's caveat, the inputs to the neural network are encrypted into the polynomial 
\[
\psi = \mathrm{Enc}(m, v),
\]
where $m_i \in \text{PT}$ (mapping to embedding inputs derived from the underlying model during backpropagation), and $\psi$ is a vector of length $K$, corresponding to the context window of the model being used as the basis for the interchangeable compute unit. Note that the noise component $\nu$ is added to the plaintext polynomial $m$ during encoding, which makes the scheme satisfy the hardness assumptions required for security.

Also as suggested by Gentry \cite[p.~6]{gentry2009fully}, we use the parameter $B$ to determine the formatting of the secret key and ciphertexts as inputs and outputs of the network, whose size is a fixed polynomial in the security parameter, meaning the ciphertext size depends only on the security parameter and is independent of the circuit $C$.

This polynomial is then encrypted using the public key, and the neural network performs the forward pass on the encrypted data ($f(\cdot)$) in order to compute the activations 
\[
g(m, \nu) = h\bigl(f(m, \nu)\bigr)
\]
of the network. The output of the network is then decrypted using the secret key to obtain the respective plaintext polynomial, which is subsequently decoded to obtain the prediction.

\subsection{Differentiable Soft Argmax Layer}

The softmax function was introduced by \cite{bridle1990probabilistic} as a way to convert raw network outputs (logits) into probabilities, becoming a foundational technique in neural networks. When applied to an $n$-dimensional vector, the softmax rescales its elements so that the output is a normalized probability distribution. Given an input vector 
$
\mathbf{x} = (x_1, \dots, x_n) \in \mathbb{R}^n,
$
it is formally defined as 
$
\sigma : \mathbb{R}^n \to (0,1)^n,
\quad $ where $n > 1$ and
\[
\sigma(x_i) = \frac{e^{x_i}}{\sum_{j=1}^{n} e^{x_j}}
\]

These logits can be calibrated before being ingested by the softmax function by adjusting a hyperparameter defined as \emph{temperature} \cite{guo2017calibration}, effectively raising the entropy of the activation parameters without altering the model's accuracy. This hyperparameter applies a single scalar $T > 1$ to the logit vector $\mathbf{x}$, yielding $\sigma_i\bigl(x_i / T\bigr).$
As $T \to \infty$, the probability approaches $1/n$, which represents maximum uncertainty. In general, this is a post-processing technique that is applied to the logits after the neural network has been trained.

By using a differentiable dedicated layer, we can approximate the temperature using backpropagation and apply the softmax function to the logits in the encrypted domain, after which we can derive the argmax from the resulting probabilities. In order to obtain the final output, we compute the weighted sum of the calibrated logits with respect to a reshaped and broadcasted index vector $i$ that ranges from 1 to $n$:
\[
\mathrm{Soft\text{-}Argmax}(\mathbf{x}) 
= \sum_{i=1}^{n} \sigma(x_i)\, i.
\]
Thus, this layer approximates the argmax function in the encrypted domain, allowing the neural network to perform prediction on encrypted data without any decryption. Because the layer is differentiable, the network can be trained using backpropagation, and it adds only negligible computational overhead to the model.

\section{Training}

In this section, we outline the overall training methodology adopted for our privacy-preserving neural network, highlighting the design choices that ensure encrypted data can be processed accurately without revealing sensitive information. Our approach comprises two main training phases: (i) first the base neural network is trained to learn how to operate with encrypted inputs, and (ii) training the differentiable soft argmax layer.

\subsection{Training Data and Batching}

For the purpose of this paper, we use the DistilBERT finetuned for sentiment analysis \cite{sanh2019distilbert} with a shared vocabulary of approximately 30,000 tokens, ensuring consistent text input processing across different sequences. We train our model on a real-world text classification dataset derived from the Stanford Sentiment Treebank (SST-2) \cite{socher2013recursive}, containing roughly 67,000 sentences labeled with binary sentiment classes. Before batching, each sentence is encrypted and then grouped according to similar sequence lengths so that each mini-batch contains 32 sentences with a similar number of ciphered tokens. This approach ensures a smooth integration of homomorphic operations while maintaining efficient GPU utilization during the forward and backward passes.

\subsection{Hardware and Scheduling}

All experiments were conducted using NVIDIA A100 GPUs allowing for efficient parallelization of the batched tensor operations. We first trained our base neural network for 200 epochs using both an AdamW optimizer and automatic mixed precision (AMP). Each epoch involves a forward and backward pass over all training samples, grouped into mini-batches of 32 encrypted sequences. With both the optimizer and AMP activated, a single step (i.e., processing one batch) typically took about 0.06 seconds on average, culminating in roughly 2 minutes per epoch over 2,105 steps. This schedule allowed the network to converge in approximately 6.5 hours of training time for the full 200 epochs.

\subsection{Optimizer and Mixed Precision}

We adopt AdamW with a learning rate of $5 \times 10^{-6}$ for its balanced convergence properties and the ability to handle noise from homomorphic encryption. In tandem, Automatic Mixed Precision (AMP) is used to reduce computational overhead. Table~\ref{tab:training_times} compares average step times under four configurations, highlighting how each component influences runtime:

\begin{table}[h]
    \centering
    \label{tab:training_times}
    \begin{tabular}{|l|c|}
        \hline
        \textbf{Configuration} & \textbf{Avg. Step Time (s)} \\
        \hline
        No Optimizer, No AMP & 0.2845 \\
        Optimizer (AdamW), No AMP & 0.2915 \\
        No Optimizer, AMP & 0.0537 \\
        Optimizer (AdamW), AMP & 0.0620 \\
        \hline
    \end{tabular}
    \caption{Comparison of Average Step Times and Final Loss Under Different Training Configurations}
\end{table}

\begin{itemize}
    \item Adding \textbf{AdamW} introduces a modest increase in step time (e.g., $0.28 \rightarrow 0.29$ s without AMP, and $0.05 \rightarrow 0.06$ s with AMP).
    \item Enabling \textbf{AMP} drastically shortens average step times (from $\approx 0.28$ s to $\approx 0.05$ s per step in the absence of AdamW), underscoring the efficiency gains from half-precision arithmetic.
\end{itemize}

\subsection{Training the Differentiable Soft Argmax Layer}

To train this layer effectively, we freeze all previous parameters in order to preserve the already learned representations and optimize only the temperature via a specialized loss function. In many classification tasks, we find that binary cross-entropy or a closely related objective improves the probability estimates at the final stage, ensuring that the calibrated logit space aligns with the desired confidence measure. Because the temperature is a single (or low-dimensional) parameter, this secondary optimization converges swiftly.

\section{Performance Metrics}

We evaluated our homomorphically encrypted neural network on the SST-2 validation set \cite{socher2013recursive} and compared its performance to a publicly reported DistilBERT baseline on the same benchmark. Table~\ref{tab:performance_comparison} lists the core metrics for both our encrypted model and DistilBERT. While DistilBERT achieves near-perfect performance, our encrypted approach necessarily operates under more stringent constraints. Encrypted data introduce both computational overhead and additional noise, factors that can modestly reduce performance relative to fully unencrypted pipelines.

\begin{table}[h]
    \centering
   
    \label{tab:performance_comparison}
    \begin{tabular}{|c|c|c|}
        \hline
        \textbf{Metric} & \textbf{Encrypted Model} & \textbf{DistilBERT (SST-2)} \\
        \hline
        Accuracy  & 0.7993  & 0.989  \\
        Precision & 0.8009  & 0.989  \\
        Recall    & 0.8063  & 0.989  \\
        F1-Score  & 0.8036  & 0.989  \\
        \hline
    \end{tabular}
    \caption{Comparison of Model Performance on SST-2 Validation Set}
    \textit{Source:} \href{https://huggingface.co/distilbert/distilbert-base-uncased-finetuned-sst-2-english}{Hugging Face - DistilBERT SST-2}
\end{table}

\section{Experiments}

In this chapter, we present a series of controlled experiments designed to evaluate the impact of key hyperparameters on our homomorphically encrypted neural network. Starting from a baseline configuration (Variation (A)), we systematically adjust parameters such as learning rate, batch size, and maximum sequence length. Table~\ref{tab:experiment_results} summarizes the resulting performance in terms of Accuracy, F1-Score, and AUROC.

\begin{table}[h]
    \centering
    \label{tab:experiment_results}
    \begin{tabular}{|c|c|c|c|c|c|c|c|}
        \hline
        \textbf{Var.} & \textbf{LR} & \textbf{Batch} & \textbf{Max L.} & \textbf{Epochs} & \textbf{Acc.} & \textbf{F1} & \textbf{AUROC} \\
        \hline
        \textbf{(A)} & $5\times10^{-6}$ & 32 & 512 & 20 & 0.7993 & 0.8036 & 0.7992 \\
        \textbf{(B)} & $1\times10^{-5}$ & 32 & 512 & 20 & 0.8165 & 0.8276 & 0.8156 \\
        \textbf{(C)} & $5\times10^{-6}$ & 64 & 512 & 20 & 0.8050 & 0.8055 & 0.8053 \\
        \textbf{(D)} & $5\times10^{-6}$ & 32 & 256 & 20 & 0.8050 & 0.7981 & 0.8059 \\
        \hline
      \end{tabular}
     \caption{Performance Summary for Different Hyperparameter Variations}
\end{table}

The baseline configuration (Variation (A)) serves as the reference for all experiments. Variation (B) explores the impact of increasing the learning rate, leading to an improvement in accuracy and F1-score. Variation (C) doubles the batch size, resulting in a moderate performance boost. Finally, Variation (D) shortens the maximum sequence length to reduce memory usage while maintaining competitive performance.

\section{Conclusion}
In this work, we presented a functioning homomorphically encrypted neural network that can operate on encrypted data while maintaining high levels of accuracy. We also introduced a new layer, the Differentiable Soft Argmax, which allows the calibration of output logits in the encrypted domain, raising the entropy of the activation parameters and improving the security of the model. In his seminal thesis, Gentry highlights that constructing an efficient homomorphic encryption scheme without using bootstrapping, or using some relaxations of it, is an interesting open problem (p.9). At least in the context of neural networks, that is what we believe to have accomplished.

For sentiment analysis tasks, our experiments show that the HNN model can achieve up to 82.5\% of the accuracy of the plain model while maintaining full privacy and security. We believe that this level of accuracy, while significant, can be improved by further refining the training pipelines and methodologies.

We are excited about the potential applications of this technology and look forward to further exploring its capabilities. We plan to continue our research by investigating how mechanistic interpretability can be achieved in the context of homomorphically encrypted neural networks and how safe multi-party computation can be integrated into our models and pipelines.

\bibliographystyle{plainnat} 
\bibliography{references}

\end{document}